# Xampling: Signal Acquisition and Processing in Union of Subspaces

Moshe Mishali, *Student Member, IEEE*, Yonina C. Eldar, *Senior Member, IEEE* and Asaf Elron

*Abstract*—We introduce Xampling, a unified framework for signal acquisition and processing of signals in a union of subspaces. The main functions of this framework are two. Analog compression that narrows down the input bandwidth prior to sampling with commercial devices. A nonlinear algorithm then detects the input subspace prior to conventional signal processing. A representative union model of spectrally-sparse signals serves as a test-case to study these Xampling functions. We adopt three metrics for the choice of analog compression: robustness to model mismatch, required hardware accuracy and software complexities. We conduct a comprehensive comparison between two sub-Nyquist acquisition strategies for spectrally-sparse signals, the random demodulator and the modulated wideband converter (MWC), in terms of these metrics and draw operative conclusions regarding the choice of analog compression. We then address lowrate signal processing and develop an algorithm for that purpose that enables convenient signal processing at sub-Nyquist rates from samples obtained by the MWC. We conclude by showing that a variety of other sampling approaches for different union classes fit nicely into our framework.

*Index Terms*—Analog to digital conversion, baseband processing, compressed sensing, digital signal processing, modulated wideband converter, sub-Nyquist, Xampling.

## I. INTRODUCTION

SIGNAL processing methods have changed substantially over the last several decades. The number of operations that are shifted from analog to digital is constantly increasing, leaving amplifications and fine tunings to the traditional front-end. Sampling theory, the gate to the digital world, is the key enabling this revolution. Traditional sampling theorems assume that the input lies in a predefined subspace [1], [2]. The most prevalent example is bandlimited sampling, according to the theorem of Shannon-Nyquist [3], [4]. Recently, nonlinear union of subspaces (UoS) models have been receiving growing interest in the context of analog sampling [5]–[12]. The UoS setting captures uncertainty in the signal by allowing several possible subspace descriptions, with the exact signal subspace unknown a-priori.

In contrast to classic subspace sampling, the theory of sampling over UoS is still developing. In particular, to date, there is no equivalent to the oblique projection operator which

This work has been submitted to the IEEE for possible publication. Copyright may be transferred without notice, after which this version may no longer be accessible.

The authors are with the Technion—Israel Institute of Technology, Haifa 32000, Israel (email: moshiko@tx.technion.ac.il; yonina@ee.technion.ac.il; elron@tx.technion.ac.il).

M. Mishali is supported by the Adams Fellowship Program of the Israel Academy of Sciences and Humanities. Y. C. Eldar is currently on leave at Stanford, USA. Her work was supported in part by the Israel Science Foundation under Grant no. 1081/07 and by the European Commission in the framework of the FP7 Network of Excellence in Wireless COMmunications NEWCOM++ (contract no. 216715).

**TABLE I:** Abbreviations Used Throughout The Paper

| ADC | analog to digital conversion |
|---|---|
| Back-DSP | Backward-compatible digital signal processing |
| CS | compressed sensing |
| CTF | continuous to finite |
| DAC | digital to analog conversion |
| DFT | discrete Fourier transform |
| DSP | digital signal processing |
| MWC | modulated wideband converter |
| RD | random demodulator |
| RF | radio frequency |
| UoS | union of subspaces |
| X-ADC | lowrate analog to digital conversion |
| X-DSP | lowrate digital signal processing |

reconstructs the signal when its exact subspace is known for almost all sampling functions [1], [2], [13]–[15]. The lack of a complete theory has not withheld development of numerous stylized applications [6]–[11], [15]–[19], aiming at reducing the sampling rate below Nyquist by exploiting the UoS model. The acquisition and reconstruction methods of [6]–[11], [15]–[19] are substantially different from each other, raising the question of whether the apparent distinct approaches can be derived from a common framework.

The first and main contribution of this paper is a unified and pragmatic framework for acquisition and processing of UoS signal classes, referred to as Xampling. It consists of two main functions: lowrate analog to digital conversion (X-ADC), in which the input is compressed in the analog domain prior to sampling with commercial devices, and lowrate digital signal processing (X-DSP), in which the input subspace is detected prior to signal processing in the digital domain. In both cases the X prefix hints at the rate reduction. After presenting the architecture in Section II, we show that a wide range of UoS applications [6]–[11], [15]–[19] fit elegantly into the proposed sampling structure.

We next study the X-ADC block and address the choice of analog compression. We do that by examining the compression techniques used in the random demodulator (RD) [19] and modulated wideband converter (MWC) [8] systems. These methods apply compressed sensing (CS) ideas to reduce the sampling rate of spectrally-sparse signals below the Nyquist rate. At first sight, the signal models and compression techniques used seem similar, at least visually. By examining three design metrics: robustness to model mismatch, required hardware accuracy and computational loads, we reveal several advantages of the MWC in all three metrics. Based on the insights gained, we draw operative guidelines for the choice of analog compression in Xampling systems that rely on CS principles. Besides the main interest in studying X-ADC, this contribution is also the first comprehensive technical comparison between the RD and MWC systems, which reveals major differences, that are not evident from the original publications [8], [19].

**TABLE II:** Applications of Union of Subspaces

| Union model | $\|\Lambda\|$ | $\dim(\mathcal{A}_\lambda)$ | Analog compression | | Subspace detection |
|---|---|---|---|---|---|
| Multiband | finite | $\infty$ | Periodic nonuniform sampling [16]: | time shifts | CTF + CS |
| | | | Modulated wideband converter [8]: | periodic mixing + lowpass | CTF + CS |
| | | | Nyquist-folding receiver [20]: | jittered undersampling (nonlinear) | n/a |
| Harmonic tones | finite | finite | Random demodulator [19]: | sign flipping + integration | CS |
| Time-delay innovation (FRI) | $\infty$ | finite | periodic [6], [21]: | lowpass | annihilating filter [6], [21] |
| | | | one-shot [22]: | splines | moments factoring [22] |
| | | | periodic/one-shot [10], [11]: | Sum-of-Sincs filtering | annihilating filter |
| Sequences of innovation | $\infty$ | $\infty$ | [9], [10]: | lowpass, or periodic mixing + integration | MUSIC [23] or ESPRIT [24] |
| Sparse shift-invariant | finite | $\infty$ | [15]: | filter-bank | CTF + CS |
| Abstract unions | finite | finite | [7]: | projection onto Riesz bases | CS |

As a third contribution, we study the X-DSP stage and sub-Nyquist processing, which is challenging since conventional DSP methods assume their input data stream is given at the Nyquist rate. We develop a digital algorithm, named Back-DSP, that provides the MWC with a smooth interface to existing DSP software. Our algorithm consists of several lowrate processing steps, which together detect the exact signal subspace, thereby gaining backward compatibility to conventional processing methods. The alternative approach of [25], which suggests the development of processing methods tailored to CS measurements, is discussed and compared to. As a nice feature, we show that once the Back-DSP algorithm is applied, the input can be reconstructed more efficiently than the original method of [8]. Numerical simulations demonstrate backward compatibility in typical noisy wideband scenarios.

The paper is organized as follows. Section II describes the UoS model and presents the Xampling framework. X-ADC and X-DSP are studied in the next two sections. The choice of analog compression is studied in Section III based on a comparison between the RD and MWC architectures. Following, in Section IV, we develop and simulate the Back-DSP algorithm. Table I lists abbreviations that are used throughout.

## II. XAMPLING

In this section, we describe the class of UoS signals and present Xampling – our proposed framework for acquisition and digital processing of these signal models.

### A. Union of Subspaces

Let $x(t)$ be an input signal in the Hilbert space $\mathcal{H} = L_2(\mathbb{R})$. The signal $x(t)$ is assumed to lie in a UoS of $\mathcal{H}$, namely within a parameterized family of subspaces

$$x(t) \in \mathcal{U} \triangleq \bigcup_{\lambda \in \Lambda} \mathcal{A}_\lambda, \tag{1}$$

where $\Lambda$ is a list of indices, and each individual subspace $\mathcal{A}_\lambda \in \mathcal{H}$. The key property of the UoS model is that the input $x(t)$ resides within $\mathcal{A}_{\lambda^*}$ for some $\lambda^* \in \Lambda$, but a-priori, the exact subspace index $\lambda^*$ is unknown. This model was originally introduced by Lu and Do in [5]. In general, the sum (or a linear combination) of $x_1(t), x_2(t) \in \mathcal{U}$ does not lie in $\mathcal{U}$. Thus, (1) typically represents a nonlinear set of possible inputs, which is a true subset of the linear sum of all subspaces $\mathcal{A}_\lambda$, denoted hereafter by $\Sigma$.

The first column of Table II lists several signal classes that can be readily modeled as UoS (see also a list of applications in [5]). We consider two motivating examples from this table. A first application of (1) is multiband sampling, encountered when a communication receiver intercepts multiple radiofrequency (RF) transmissions, but is not provided with their carrier frequencies $f_i$. In this setting, the input $x(t)$ has multiband spectra with energy that concentrates on $N$ frequency intervals of individual widths $B$ located anywhere below some maximal frequency $f_{\max}$. Such a receiver faces a challenging sampling problem, since classic acquisition methods, such as RF demodulation or bandpass undersampling, require knowledge of the values $f_i$. At first sight, it may seem that sampling at the Nyquist rate

$$f_{\text{NYQ}} = 2f_{\max}, \tag{2}$$

is necessary, since every frequency interval below $f_{\max}$ can potentially contain a transmission of interest. On the other hand, since each specific $x(t)$ fills only a portion of the Nyquist range (only $NB$ Hz), one would intuitively expect to be able to reduce the sampling rate below $f_{\text{NYQ}}$.

A multiband model can be described in union terminology by indexing the possible band positions with $\lambda = \{f_i\}$ and letting $\mathcal{A}_\lambda$ capture the subspace of multiband signals on the chosen support. It is therefore expected that an input from a multiband union can be determined from sampling at a rate proportional to the actual bandwidth occupied by $\mathcal{A}_\lambda$, namely $NB$, up to some rate increase needed to determine the unknown subspace index $\lambda^*$ of the given $x(t)$. In principle, $f_i$ lies in the continuum $f_i \in [0, f_{\max}]$ in this modeling, so that the union contains infinitely many subspaces. A different viewpoint, utilized in [8], [16], is to divide the Nyquist range to $M$ slices and enumerate the possible supports according to the slice indices that contain signal energy. This approach results in a finite union of bandpass subspaces, which enables efficient hardware and software implementation, as further discussed in detail in Section III-B.

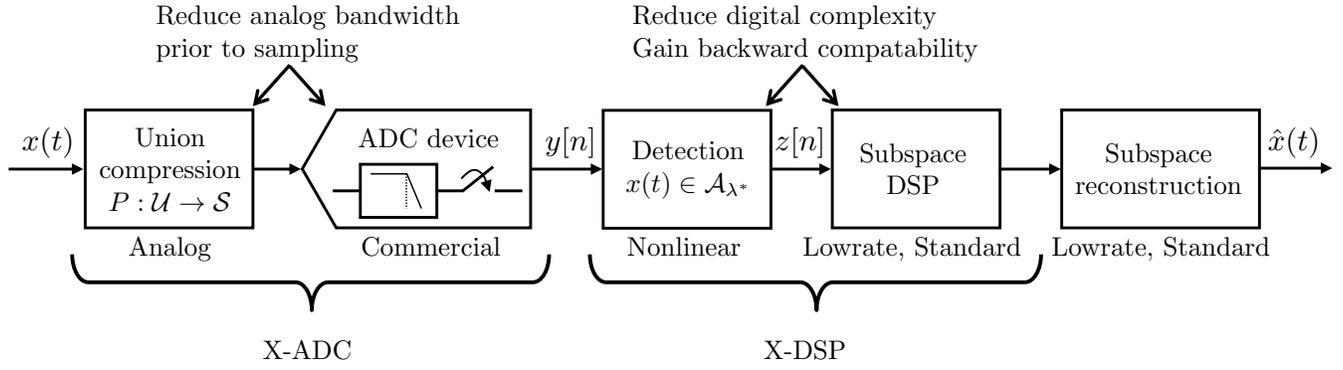

**Fig. 1:** Xampling – A pragmatic framework for signal acquisition and processing in union of subspaces.

Another interesting application is estimation of time delays from observation of a signal of the following form

$$x(t) = \sum_{\ell=1}^{L} d_\ell \, g(t - t_\ell), \quad t \in [0, T]. \qquad (3)$$

Inputs of this type belong to a broader family of signals with a finite rate of innovation (FRI) [6], [22]. In practice, there are many interesting situations with unknown $t_\ell$, which can be modeled in union terminology by assigning the delays $\lambda = \{t_\ell\}$ for the index of an $L$-dimensional subspace $\mathcal{A}_\lambda$ of FRI signals, spanned by the amplitudes $a_\ell$. Such an FRI union is encountered, for example, when a channel with multipath fading generates echoes of a transmitted pulse $g(t)$ in various unknown delays and attenuations [9], or in radar [12], where $t_\ell$ and $a_\ell$ correspond to target locations and speeds, respectively. Ultrasound imaging [11] and underwater acoustics also conform with (3). Since in all these applications, the pulse $g(t)$ is short in time, sampling $x(t)$ according to its Nyquist bandwidth, which is effectively that of $g(t)$, results in unnecessary large sampling rates. Classic match filtering methods require Nyquist rate sampling [9]. In contrast, union modeling implies a rate requirement that is proportional to the innovation rate $2L/T$, which in all the above applications can be substantially lower than Nyquist.

### B. Unified Goals

The above examples imply that treating UoS models at low rates calls for sophisticated acquisition and processing methods in order to exploit the underlying structure. In principle, one could employ traditional techniques developed in sampling theory for linear single-subspace scenarios [1], [2], by sampling $\Sigma$, namely the linear sum of all subspaces $\mathcal{A}_\lambda$. However, this technically-correct approach often leads to practically-infeasible sampling systems wasting expensive hardware and software resources. For example, in multiband sampling, $\Sigma$ is the $f_{\max}$-bandlimited space, for which no rate reduction is possible. Similarly, in time-delay estimation problems, $\Sigma$ has the high bandwidth of $g(t)$, and again no rate reduction can be achieved. To benefit from the union structure, we need to incorporate its nonlinear structure and exploit the fact that $\mathcal{U}$ is typically a true subset of $\Sigma$.

To be a bit more precise, we define the sampling problem for the union set (1) as the design of a system that provides:

1) **ADC:** an acquisition operator which converts the analog input $x(t) \in \mathcal{U}$ to a sequence $y[n]$ of measurements,
2) **DSP:** a toolbox of processing algorithms, which uses $y[n]$ to perform classic tasks, *e.g.,* estimation, detection, data retrieval etc., and
3) **DAC:** a method for reconstructing $x(t)$ from the samples $y[n]$.

In order to exclude from consideration inefficient solutions, such as those treating the Nyquist subspace $\Sigma$ and not exploiting the union structure, we adopt as a general design constraint that the above goals should be accomplished with minimum use of resources. Minimizing the sampling rate, for example, excludes inefficient Nyquist-rate solutions and tunnel potential approaches to wisely incorporate the union structure to stand this resource constraint. For reference, this requirement is outlined as

$$\textbf{ADC + DSP + DAC} \rightarrow \textbf{minimum use of resources.} \quad (4)$$

In practice, besides constraining the sampling rate, (4) translates to the minimization of several other resources of interest, including the number of devices in the acquisition stage, design complexity, processing speed, memory requirements, power dissipation, system cost, and more. As we shall see via examples in the sequel, the challenge posed in (4) is to treat a union model at an overall complexity (of hardware and software) that is comparable with a system which knows the exact $\mathcal{A}_{\lambda^*}$.

As evident from Table II, different instances of UoS models have received treatment using quite different hardware and software techniques. In the next subsection we introduce Xampling, our proposed architecture to unify the sampling of UoS signal classes and address the resource constraint (4).

### C. Architecture

The Xampling framework we propose has the high-level architecture presented in Fig. 1. As highlighted, the Xampling architecture is driven by two main considerations: reducing analog bandwidth prior to sampling and gaining lowrate DSP, preferably backward-compatible with existing processing algorithms. We next describe the five functional blocks and explain these two considerations.

**Analog bandwidth compression.** The first two blocks, termed X-ADC, perform the conversion of $x(t)$ to digital. An operator $P$ compresses the high-bandwidth input $x(t)$ into a signal with lower bandwidth, effectively capturing the entire union $\mathcal{U}$ by a subspace $\mathcal{S}$ with substantially lower sampling requirements. A commercial ADC device then takes pointwise samples of the compressed signal, resulting in the sequence of samples $y[n]$.

The role of $P$ in Xampling is to narrow down the analog bandwidth that enters the acquisition devices, so that lowrate ADC devices can be used. Actual acquisition is modeled in Fig. 1 as a lowpass filter followed by a pointwise sampler, with the lowpass reflecting a limited front-end bandwidth of the conversion device. The X-ADC can be realized on a circuit board, chip design, optical system or other appropriate hardware. In all these platforms, the front-end has certain bandwidth limitations, which stem from the responses of all circuitries comprising the internal front-end. Commercial ADC devices are often specified with front-end bandwidth that is not much wider than twice their sampling rate capabilities [8]. Thus, direct acquisition of pointwise values of $x(t)$ with commercial components generally requires an ADC device with Nyquist-rate bandwidth, even when taking pointwise values at a low rate. In Xampling, the input signal $x(t)$ belongs to a union set $\mathcal{U}$ which typically has high bandwidth, *e.g.*, multiband signals whose spectrum reaches up to $f_{\max}$ or FRI signals with wideband pulse $g(t)$. A preceding analog compression step $P$ is therefore necessary in order to capture all vital information within a narrow range of frequencies.

In contrast to popular compression techniques that are realized in software, here $P$ captures all vital information of the input by hardware preprocessing. The design of $P$ therefore needs to properly exploit the union structure, in order not to lose any essential information while reducing the bandwidth. The next stage can then employ commercial devices with low analog bandwidth, as part of minimizing resource usage (4).

**Lowrate DSP.** A second goal of Xampling is to translate the sampling rate reduction to a comparable decrease in processing speeds. Our proposal for achieving this goal consists of the three computational blocks in the digital part of Fig. 1. A nonlinear step detects the signal subspace $\mathcal{A}_{\lambda^*}$ from the lowrate samples. Once the index $\lambda^*$ is determined, we compute a low-rate sequence $z[n]$ of numbers that matches standard sampling of $\mathcal{A}_{\lambda}$. As a nice feature, this creates a seamless interface to existing DSP algorithms and interpolation techniques, hence provides backward compatibility. The combination of nonlinear detection and standard DSP is referred to as X-DSP. Besides backward compatibility, the nonlinear detection decreases computational loads, since the subsequent DSP and DAC stages need to treat only the single subspace $\mathcal{A}_{\lambda^*}$, complying with (4). The important point is to detect $\lambda^*$ and compute $z[n]$ without going through reconstruction of the Nyquist-rate samples of $x(t)$, or through Nyquist-rate computations.

Lowrate DSP can sometimes be an important requirement, regardless of whether the sampling rate is reduced as well. In particular, the digital flow proposed in Fig. 1 is beneficial even when a high ADC rate is acceptable. In this case, $x(t)$ can be acquired directly without narrowing down its bandwidth prior to ADC, but we would still like to reduce computational loads and storage requirements in the digital domain. This can be accomplished by imitating rate reduction in software, detecting the signal subspace and processing at the information bandwidth. Compounded usage of both X-ADC and X-DSP is for mainstream applications, where reducing the rate of both signal acquisition and processing is of interest.

Xampling is a generic template architecture. It does not specify the exact acquisition operator $P$ or nonlinear detection method to be used. These are application-dependant functions. Our goal in introducing Xampling is to propose a high-level system architecture and a basic set of guidelines:

1) an analog pre-processing unit to compress the input bandwidth,
2) commercial lowrate ADC devices for actual acquisition at a low rate,
3) subspace detection in software, and
4) standard DSP and DAC methods.

As a first step in establishing the framework, we summarize in Table II various recent sampling strategies and identify their compression and detection blocks. It can verified that the apparent different acquisition stages aim all at capturing signal information using only a small set of lowrate sample sequences, and that in all scenarios, the digital algorithms determine the input subspace as part of reconstruction. This affirms that Xampling is sufficiently general to capture a variety of UoS applications in a unified manner.

In the next two sections, we consider a representative union model of spectrally-sparse signals in order to study in more detail practical considerations in designing the analog compression stage $P$ and subspace detection algorithms that provide lowrate DSP.

## III. X-ADC: Sub-Nyquist Signal Acquisition

In this section, we study the X-ADC stage of Fig. 1 and in particular the analog compression operator $P$. This stage needs to be realized in hardware as it precedes the sampler. In practice, hardware imperfections are inevitable and real-world inputs may not perfectly fit the theoretical signal model. Therefore, two metrics of interest in choosing $P$ are robustness to model mismatch and required hardware accuracy. In addition, since $P$ is effectively inverted by subsequent digital recovery algorithms, the impact on computational loads is a third metric to consider.

We study the way analog compression is realized in two similar systems: RD [19] and MWC [8]. The RD treats a sparse sum of harmonic tones, whereas the MWC samples multiband signals. In both cases, analog compression involves mixing the input with certain waveforms prior to sampling, and reconstruction relies on CS techniques. Despite the seemingly-similar setup, our study reveals significant differences in terms of the three metrics we consider, and leads to several operative suggestions regarding the choice of $P$ in Xampling systems that incorporate CS principles. Besides our prime focus on X-ADC design, this study provides the first comprehensive technical comparison between these two systems.

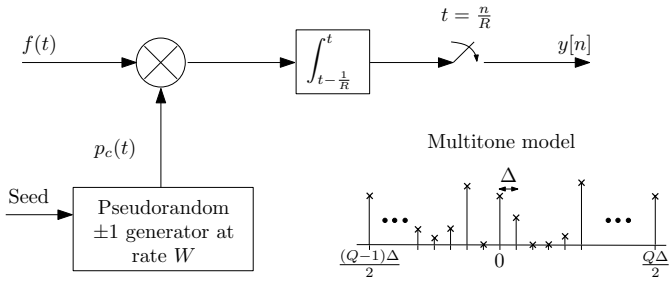

**Fig. 2:** Block diagram of the random demodulator [19].

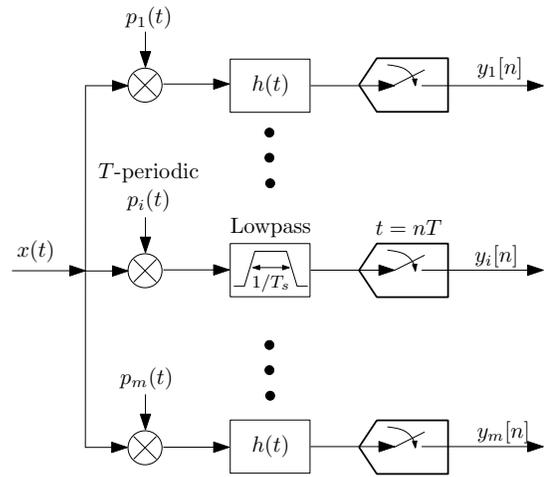

**Fig. 3:** Block diagram of the modulated wideband converter [8].

### A. Random Demodulator

The RD approach treats signals consisting of a discrete set of harmonic tones with the system that is depicted in Fig. 2.

**Signal model.** A multitone signal $f(t)$ consists of a sparse combination of integral frequencies:

$$f(t) = \sum_{\omega \in \Omega} a_\omega e^{j2\pi\omega t}, \tag{5}$$

where $\Omega$ is a finite set of $K$ out of an even number $Q$ of possible harmonics

$$\Omega \subset \{0, \pm\Delta, \pm 2\Delta, \cdots, \pm(0.5Q-1)\Delta, 0.5Q\Delta\}. \tag{6}$$

The model parameters are the tone spacing $\Delta$, number of active tones $K$ and grid length $Q$. The Nyquist rate is $Q\Delta$. In [19], the tones spacing is normalized to $\Delta = 1$ Hz. To this end, whenever normalized, $\Delta$ is omitted from formulas under the convention that all variables take nominal values (*e.g.,* $R = 10$ instead of $R = 10$ Hz).

**Sampling.** The input signal $f(t)$ is mixed by a pseudorandom chipping sequence $p_c(t)$ which alternates at a rate of $W$. The mixed output is then integrated and dumped at a constant rate $R$, resulting in the sample sequence $y[n]$, $1 \le n \le R$. The development in [19] uses the following parameter setup

$$\Delta = 1, \quad W = Q, \quad R \in \mathbb{Z} \text{ such that } \frac{W}{R} \in \mathbb{Z}. \tag{7}$$

It was proven in [19] that if $W/R$ is an integer and (7) holds, then the vector of samples $\mathbf{y} = [y[1], \ldots, y[R]]^T$ can be written as

$$\mathbf{y} = \mathbf{\Phi}\mathbf{x}, \quad \mathbf{x} = \mathbf{F}\mathbf{s}, \quad \|\mathbf{s}\|_0 \le K. \tag{8}$$

The matrix $\mathbf{\Phi}$ has dimensions $R \times W$, effectively capturing the mechanism of integration over $W/R$ Nyquist intervals, where the polarity of the input is flipped on each interval according to the chipping function $p_c(t)$. See Fig. 6(a) in the sequel for further details on $\mathbf{\Phi}$. The $W$-squared discrete Fourier transform (DFT) matrix $\mathbf{F}$ accounts for the sparsity in the frequency domain. The vector $\mathbf{s}$ has $Q$ entries $s_\omega$ which are up to a constant scaling from the corresponding tone amplitudes

$a_\omega$. Since the signal has only $K$ active tones, $\|\mathbf{s}\|_0 \le K$, where the $\ell_0$-norm counts the number of nonzero entries.

**Reconstruction.** The unknown in (8) is $\mathbf{s}$. Observe that $\mathbf{y} = \mathbf{\Phi}\mathbf{F}\mathbf{s}$ does not determine $\mathbf{s}$ by itself, since $\mathbf{\Phi}\mathbf{F}$ is underdetermined, *i.e.,* has less rows than columns, $R < W$. An underdetermined system has a nontrivial null space and infinitely many solutions in general. Among these solutions, (8) requires the one with $\|\mathbf{s}\|_0 \le K$. This type of problem has received extensive treatment in the CS literature, where $\mathbf{\Phi}$ is referred to as the sensing matrix and $\mathbf{F}$ is termed the sparsity basis of $\mathbf{x}$. Under mild conditions on $\mathbf{\Phi}\mathbf{F}$, (8) has a unique sparse solution $\mathbf{s}$ [17], [18]. Whilst finding a sparse solution is NP-hard in general, several polynomial-time CS techniques are known to coincide with the true $\mathbf{s}$ under certain conditions on $\mathbf{\Phi}\mathbf{F}$. Example techniques include $\ell_1$ minimization, *a.k.a.,* basis pursuit [26], and greedy-type algorithms; cf. [27]. Roughly speaking, we say that $\mathbf{\Phi}$ is a "nice" CS matrix, if (8) with sparsity order $K$ can be solved efficiently with existing polynomial-time algorithms[1]. Correct recovery with a "nice" $\mathbf{\Phi}$ requires a sampling rate on the order of [19]

$$R \approx 1.7K \log(W/K + 1). \tag{9}$$

Once the sparse $\mathbf{s}$ is found, the amplitudes $a_\omega$ are determined from $s_\omega$ by constant scaling, and the output $\hat{f}(t)$ is synthesized according to (5).

### B. Modulated Wideband Converter

The MWC system samples multiband signals with the system that is depicted in Fig. 3.

**Signal model.** A multiband signal $x(t)$ has sparse spectra, supported on $N$ frequency bands, with individual widths not exceeding $B$ Hz. The band positions are anywhere below $f_{\max}$. Fig. 4 illustrates a typical multiband spectra.

**Sampling.** The input $x(t)$ passes through RF processing front-end of $m$ channels. In the $i$th channel, $x(t)$ is multiplied by a periodic waveform $p_i(t)$ with period $T$, lowpass filtered

---

[1] We comment that most known constructions of "nice" CS matrices involve randomness. In practice, $p_c(t)$ is fixed and defines a deterministic $\mathbf{\Phi}$.

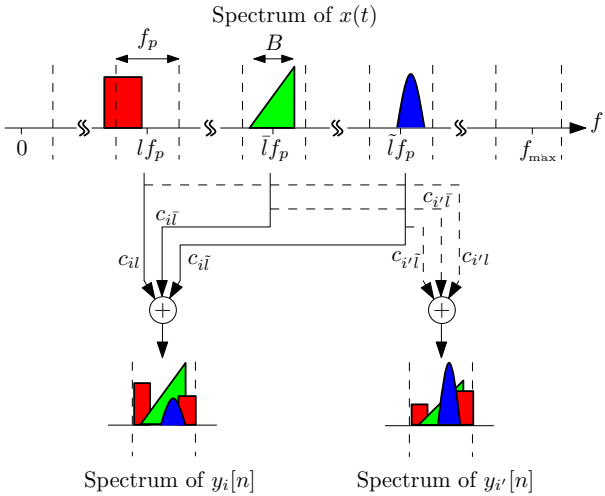

**Fig. 4:** Spectrum slices of $x(t)$ are overlayed in the spectrum of the output sequences $y_i[n]$ (taken from [28]). Since $1/T \geq B$, a single band occupies at most 2 adjacent spectrum slices. In the example, channels $i$ and $i'$ realize different linear combinations of the spectrum slices centered around $lf_p, \bar{l}f_p, \tilde{l}f_p$. For simplicity, the aliasing of the negative frequencies is not drawn.

by $h(t)$ with cutoff $1/2T$, and then sampled at rate $f_s = 1/T$. The basic parameter setting is [8]

$$m \geq 4N, \quad f_s = \frac{1}{T} \geq B. \quad (10)$$

An advanced configuration enables to collapse the number of branches $m$ by a factor of $q$ at the expense of increasing the sampling rate of each channel by the same factor, so that $f_s = q/T$. The overall sampling rate $mf_s$ is unchanged [8]. In principle, the MWC system can be collapsed to a single sampling branch using $q = m$. For the purpose of studying X-ADC, the basic version with $q = 1$ is analyzed.

Since $p_i(t)$ is periodic, it has a Fourier expansion

$$p_i(t) = \sum_{l=-\infty}^{\infty} c_{il} e^{j\frac{2\pi}{T}lt}, \quad (11)$$

with spectra that consists of a weighted Dirac-comb, with Dirac locations on $f = lf_p$ and weights $c_{il}$, where $f_p = 1/T$. Denote by $z_l[n]$ the sequence that would have been obtained if the signal was mixed by a pure sinusoid $e^{j2\pi lt/T}$ and lowpass filtered, so that $z_l[n]$ are samples of the contents in a width-$f_p$ slice of the spectrum around $lf_p$ Hz. The input $x(t)$ is determined by $z_l[n], -L \leq l \leq L$, where $L$ is the smallest index such that $Lf_p \geq f_{\max}$. Together, $M = 2L + 1$ such spectral slices cover the entire Nyquist range $[-f_{\max}, f_{\max}]$. Choosing $f_p \geq B$ ensures that a single band occupies at most 2 adjacent spectrum slices; see Fig. 4. Under this choice, the vector of samples $\mathbf{y}[n] = [y_1[n], \ldots, y_m[n]]^T$ obtained at time instant $t = nT$ satisfies the underdetermined system

$$\mathbf{y}[n] = \mathbf{C}\mathbf{z}[n], \quad \|\mathbf{z}[n]\|_0 \leq 2N, \quad (12)$$

with $\mathbf{C}$ an $m \times M$ matrix whose entries are $c_{il}$, and $\mathbf{z}[n] = [z_{-L}[n], \ldots, z_0[n], \ldots, z_L[n]]^T$. Conceptually, the MWC shifts a weighted-sum of these slices to the origin, with the lowpass filter $h(t)$ transferring only the narrow band frequencies up to $f_s/2$ from that sum to the output sequence $y_i[n]$ [8]. This aliasing structure is illustrated in Fig. 4.

The periodic functions $p_i(t)$ define the sensing matrix $\mathbf{C}$ in (12) through their Fourier coefficients $c_{il}$. Thus, $p_i(t)$ need to be chosen such that the resulting $\mathbf{C}$ has "nice" CS properties. In principle, any periodic function with high-speed transitions within the period $T$ can satisfy this requirement. One possible choice for $p_i(t)$ is a sign-alternating function, with $M = 2L + 1$ sign intervals within the period $T$ [8]. Popular binary patterns, *e.g.,* Gold or Kasami sequences, are especially suitable for the MWC [29].

**Reconstruction.** In principle, we can solve for the sparsest solution $\mathbf{z}[n]$ of (12) for every $n$, and then reconstruct $x(t)$ by properly re-positioning the slices on the spectrum. A more efficient approach, termed continuous to finite (CTF) [16], [30], exploits the fact that $\mathbf{z}[n]$ are jointly sparse over time, so that the index set $\lambda = \{l \, | \, z_l[n] \neq 0\}$ is constant over consecutive time instances $n$. The CTF recovers $\lambda$ as follows. First, it constructs a matrix $\mathbf{V}$ from several (typically $2N$) consecutive samples $\mathbf{y}[n]$, either by directly stacking $\mathbf{y}[n]$ into the columns of $\mathbf{V}$, or via other simple computations that allow combating noise [8]. Then, it solves the following underdetermined system (which is independent of $n$):

$$\mathbf{V} = \mathbf{C}\mathbf{U}, \quad \|\mathbf{U}\|_0 \leq 2N, \quad (13)$$

with $\|\mathbf{U}\|_0$ counting the number of nonidentically-zero rows in $\mathbf{U}$. It is proved in [16], that $\mathbf{U}$ has nonzero rows in locations that coincide with the indices in $\lambda$.

Once $\lambda$ is found, (12) reduces to $\mathbf{y}[n] = \mathbf{C}_\lambda \mathbf{z}[n]$, with $\mathbf{C}_\lambda$ being the appropriate column subset of $\mathbf{C}$. Pseudo-inversion of $\mathbf{C}_\lambda$ enables real-time reconstruction from that point on; one matrix-vector multiplication per incoming vector of samples $\mathbf{y}[n]$ recovers

$$\mathbf{z}_\lambda[n] = \mathbf{C}_\lambda^\dagger \mathbf{y}[n] = (\mathbf{C}_\lambda^H \mathbf{C}_\lambda)^{-1} \mathbf{C}_\lambda^H \mathbf{y}[n], \quad (14)$$

where $(\cdot)^H$ denotes hermitian conjugate and $\mathbf{z}_\lambda[n]$ are the entries of $\mathbf{z}[n]$ indicated by $\lambda$. Standard DAC techniques reconstruct $\hat{x}(t)$ via lowpass interpolation of $z_l[n], l \in \lambda$ and modulation to the proper positions on the spectrum. Polynomial-time solvers for (13) were developed in [7], [27], [30]–[36]. The required sampling rate is on the order of [8]

$$mf_s \approx 4NB \log(M/2N + 1). \quad (15)$$

We note that multiband signals with time-varying carriers can be treated by re-initiating the CTF procedure upon detection of a spectral change. Further details and simulations with time-varying multiband inputs appear in [8].

At first sight, the RD and MWC technologies seem similar, at least in their sampling stages, which involve mixing followed by either integration in Fig. 2 or lowpass filtering in Fig. 3. The difference is in the details which we study below.

### C. Comparison – Robustness to Model Mismatch

The RD system is sensitive to inputs with tones slightly displaced from the theoretical grid. To see this, we repeat the developments of [19] for an unnormalized multitone model, with $\Delta$ as a free parameter and $W, R$ that are not necessarily

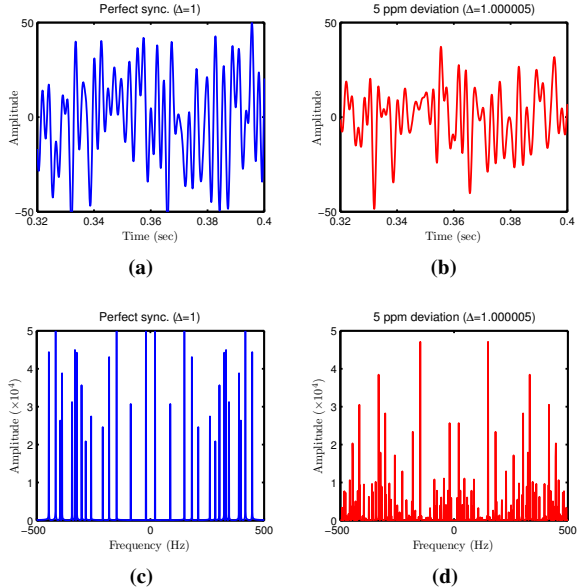

**Fig. 5:** Effects of non-integral tones on the output of the random demodulator. The top (bottom) panels plot the recovered signal in the time (frequency) domain.

integers. The measurements still obey the underdetermined system (8) as before, where now

$$W = Q\Delta, \quad R = N_R\Delta, \quad \frac{W}{R} \in \mathbb{Z}, \tag{16}$$

and $N_R$ is the number of samples taken by the RD. We refer to the technical report [37] for the exact derivation of (16). The equalities in (16) imply that the rates $W, R$ need to be perfectly synchronized with the tone spacing $\Delta$. If (16) does not hold, either due to hardware imperfections so that the rates $W, R$ deviate from their nominal values, or due to model mismatch so that the actual spacing $\Delta$ is different than what was assumed, then the reconstruction error grows high.

The following toy-example demonstrates this sensitivity. Let $W = 1000, R = 100$ Hz, with $\Delta = 1$ Hz. Construct $f(t)$ by drawing $K = 30$ locations uniformly at random on the tones grid and normally-distributed amplitudes $a_\omega$. In our simulation, basis pursuit gave exact recovery $\hat{f}(t) = f(t)$ for $\Delta = 1$. For 5 part-per-million (ppm) deviation in $\Delta$ the squared-error reached 37%:

$$\Delta = 1 + 0.000005 \quad \rightarrow \quad \frac{\|f(t) - \hat{f}(t)\|^2}{\|f(t)\|^2} = 37\%. \tag{17}$$

Figure 5 plots $f(t)$ and $\hat{f}(t)$ in time and frequency, revealing many spurious tones due to the model mismatch. The equality $W = Q$ in the normalized setup (7) hints at the required synchronization, though the dependency on the tones spacing is implicit since $\Delta = 1$. With $\Delta \neq 1$, this issue appears explicitly. Since the publication of the technical report [37], this problem was studied in [38] and [39], where it is referred to as nonintegral harmonics or sensitivity to basis mismatch, respectively.

The MWC is less sensitive to model mismatches in comparison. The parameters are set with inequalities in (10), so that the number of branches $m$ and aliasing rate $f_p$ can be chosen with some safeguards with respect to the specified number of bands $N$ and individual widths $B$. Thus, the system can handle inputs with more than $N$ bands and widths larger than $B$, up to the safeguards that were set. The band positions are not restricted to any specific displacement with respect to the spectrum slices; a single band can split between slices, as depicted in Fig. 4. A possible shortcoming in the MWC approach is the requirement to specify a multiband spectra by a pair of maximal quantities $(N, B)$. This modeling can be inefficient (in terms of resulting sampling rate) when the individual band widths are significantly different from each other. For example, a multiband model with $N_1$ bands of lengths $B_1 = k_1 b$ and $N_2$ bands of lengths $B_2 = k_2 b$ is described by a pair $(N_1 + N_2, \max(B_1, B_2))$, with spectral occupation potentially larger than actually used. A more flexible modeling in this scenario would assume only the total actual bandwidth being occupied, *i.e.*, $N_1 B_1 + N_2 B_2$. This issue can partially be addressed by designing an MWC system to accommodate $N_1 k_1 + N_2 k_2$ bands of lengths $b$.

### D. Comparison – Hardware Complexity

We next compare the hardware complexity of the RD/MWC systems. In both approaches, the acquisition stage is mapped to an underdetermined CS system: Fig. 2 leads to the sparse recovery problem (8) in the RD system, while in the MWC approach, Fig. 3 results in (12). A crucial point is that the hardware needs to be sufficiently accurate for that mapping to hold, since this is the key for reconstruction. While the RD and MWC sampling stages seem similar, they rely on different analog properties of the hardware to ensure accurate mapping to CS, which in turn imply different design complexities.

Figure 6 shall assist us in this discussion. The figure depicts the Nyquist-equivalent of each method, which is the system that samples the input at its Nyquist rate and then computes the relevant sub-Nyquist samples by applying the sensing matrix digitally. The RD-equivalent integrates and dumps the input at rate $W$, and then applies $\mathbf{\Phi}$ on $Q$ serial measurements, $\mathbf{x} = [x[1], \cdots, x[Q]]^T$. To coincide with the sub-Nyquist samples of Fig. 2, $\mathbf{\Phi} = \mathbf{HD}$ is used, where $\mathbf{D}$ is diagonal with $\pm 1$ entries, according to the values $p_c(t)$ takes on $t = n/W$, and $\mathbf{H}$ sums over $W/R$ entries [19]. The MWC-equivalent has $M$ channels, with the $l$th channel demodulating the relevant spectrum slice to the origin and sampling at rate $1/T$, which results in $z_l[n]$. The sensing matrix $\mathbf{C}$ is applied on $\mathbf{z}[n]$. Note that 6(b) is reminiscent of analog-digital hybrid filter-bank methods that are useful in high-speed ADC systems [40], [41]. While sampling according to the equivalent systems of Fig. 6 is a clear waste of resources, it enables us to view the internal mechanism of each strategy. Note that the reconstruction algorithms remain the same; it does not matter whether the samples were actually obtained at a sub-Nyquist rate, according to Figs. 2 or 3, or if they were computed after sampling according to Fig. 6.

**Hardware accuracy.** In the RD approach, time-domain properties of the hardware dictate the necessary accuracy. For example, the impulse-response of the integrator needs to be a square waveform with width of $1/R$ seconds, so that $\mathbf{H}$ has exactly $W/R$ consecutive 1's in each row. For a diagonal $\mathbf{D}$, the sign alternations of $p_c(t)$ need to be sharply aligned on $1/W$ time intervals. If either of these properties is nonideal, then the mapping to CS becomes nonlinear and signal dependent. Precisely, (8) becomes [19]

$$\mathbf{y} = \mathbf{H}(\mathbf{x})\mathbf{D}(\mathbf{x})\mathbf{x}. \qquad (18)$$

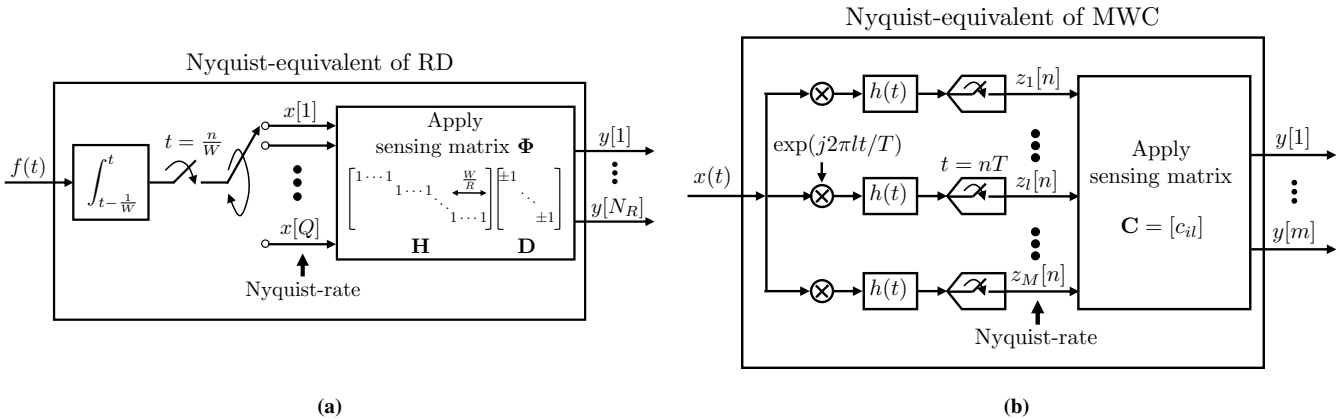

**Fig. 6:** The Nyquist-equivalents of the (a) RD and (b) MWC sample the input at its Nyquist rate and apply the sensing matrix digitally.

A noninteger ratio $W/R$ affects both $\mathbf{H}$ and $\mathbf{D}$ [19]. Since $f(t)$ is unknown, $\mathbf{x}$, $\mathbf{H}(\mathbf{x})$ and $\mathbf{D}(\mathbf{x})$ are also unknown. It is suggested in [19] to train the system on example signals, so as to approximate a linear system. Note that if (16) is not satisfied, then the DFT expansion also becomes nonlinear and signal-dependent $\mathbf{x} = \mathbf{F}(\Delta)\mathbf{s}$. The *form factor* of the RD is therefore the time-domain accuracy that can be achieved in practice.

The MWC requires periodicity of the waveforms $p_i(t)$ and lowpass response for $h(t)$, which are both frequency-domain properties. The sensing matrix $\mathbf{C}$ is constant as long as $p_i(t)$ are periodic, regardless of the time-domain appearance of these waveforms. Nonideal time-domain properties have therefore no effect on the MWC. The consequence is that stability in the frequency domain dictates the form factor of the MWC. For example, 2 GHz periodic functions were demonstrated in a circuit prototype of the MWC, where simple hardware wirings ensured that $p_i(t) = p_i(t + T)$ for every $t \in \mathbb{R}$ [28]. More broadly, circuit publications report the design of high-speed sequence generators up to 23 and even 80 GHz speeds [42], [43], where stable frequency properties are verified experimentally. Accurate time-domain appearance is not considered a design factor in [42], [43], and is in fact not maintained in practice as shown in [28], [42], [43].

The MWC scheme requires an ideal lowpass filter $h(t)$ with rectangular frequency response, which is difficult to implement due to its sharp edges. This problem appears as well in Nyquist sampling, where it is addressed by alternative sampling kernels with smoother edges at the expense of oversampling. Similar edge-free filters $h(t)$ can be used in the MWC system with slight oversampling [44]. Ripples in the passband and non-smooth transitions in the frequency response can be compensated for digitally using the algorithm in [45].

**Sampling rate.** An integer ratio $W/R$, in (7) and (16), generally requires a substantial rate increase above the theoretical rate requirement (9). The MWC does not limit the rate granularity, and in principle, can approach (15). A numerical comparison in the next subsection demonstrates this difference.

**Continuous reconstruction.** Synthesizing a multitone output $\hat{f}(t)$ requires $K$ oscillators, one per each active tone, which can be hardware excessive. Computing (5) digitally needs a processing rate of $W$, and then a DAC device at the same rate. Thus, the reconstruction complexity of the RD scales with the Nyquist rate. The MWC reconstructs $\hat{x}(t)$ using commercial DAC devices, running at the low rate $f_s = 1/T$. It needs $2N$ branches. For comparison, we note that wideband continuous inputs require prohibitively large $K, W$ to be adequately represented on a discrete grid of tones. In contrast, despite the infinitely many frequencies that comprise a multiband input, $N$ is typically small.

We note however that MWC may run into difficulties in reconstructing contents around the frequencies $(l + 0.5)f_p$, $-L \leq l \leq L$, since these are irregular points of transitions between spectrum slices. Reconstruction accuracy of these irregular points depends on the cutoff curvature of $h(t)$ and relative amplitudes of consecutive $c_{il}$. Reconstruction of an input consisting of pure tones at these specific frequencies may be imperfect. In practice, the bands encode information signals, which can be reliably decoded, even when signal energy is located around the frequencies $(l+0.5)fp$. For example, Section IV below considers multiband transmissions that carry digital information bits. We develop an algorithm that recovers the digital information bits when the noise is not too high, even when a band energy is split between adjacent slices. This algorithm also allows reconstructing $x(t)$ with only $N$ DAC devices instead of $2N$ that are required for arbitrary multiband reconstruction. Table III summarizes the model and hardware comparison.

**TABLE III:** Model and Hardware Comparison

|  | **RD** (multitone) | **MWC** (multiband) |
|---|---|---|
| Model parameters | $K, Q, \Delta$ | $N, B, f_{\max}$ |
| System parameters | $R, W, N_R$ | $m, 1/T$ |
| Setup | (7) | (10) |
|  | Sensitive, eq. (16), Fig. 5 | Robust |
| Form factor | time-domain appearance | frequency-domain stability |
| Requirements | accurate $1/R$ integration | periodic $p_i(t)$ |
|  | sharp alternations $p_c(t)$ |  |
| ADC topology | integrate-and-dump | commercial |
| Rate | gap due to (7) | approach minimal |
| DAC | 1 device at rate $W$ | $N$ devices at rate $f_s$ |

### E. Comparison – Computational Loads

In this subsection, we compare computational loads when treating multiband signals, either using the MWC system or in the RD framework by discretizing the continuous frequency axis to a grid of $Q = f_{\mathrm{NYQ}}$ tones, out of which only $K = NB$ are active [19]. We emphasize that the RD system was designed for multitone inputs, though for the study of computational loads we examine the RD on multiband inputs by considering a comparable grid of tones of the same Nyquist bandwidth. Table IV compares the RD and MWC for an input with 10 GHz Nyquist rate and 300 MHz spectral occupancy. For the RD we consider two discretization configurations, $\Delta = 1$ Hz and $\Delta = 100$ Hz. The table reveals high computational loads that stem from the dense discretization that is required to represent an analog multiband input. We also included the sampling rate and DAC speeds to complement the previous section. The notation in the table is self-explanatory, though a few aspects are emphasized below.

The sensing matrix $\boldsymbol{\Phi} = \mathbf{HD}$ of the RD has dimensions

$$\boldsymbol{\Phi} : R \times W \propto K \times Q \quad \text{(large)}. \tag{19}$$

The dimension scales with the Nyquist rate; already for $Q = 1$ MHz Nyquist-rate input, there are 1 million unknowns in (8). The sensing matrix $\mathbf{C}$ of the MWC has dimensions

$$\mathbf{C} : m \times M \propto N \times \frac{f_{\mathrm{NYQ}}}{B} \quad \text{(small)}. \tag{20}$$

For the comparable spectral occupancy we consider, $\boldsymbol{\Phi}$ has dimensions that are 6 to 8 orders of magnitude higher, in both the row and column dimensions, than the MWC sensing matrix $\mathbf{C}$. The sensing matrix size is a prominent factor since it affects many digital complexities: the delay and memory length associated with collecting the measurements, the number of multiplications when applying the sensing matrix on a vector and the storage requirement of the matrix. See the table for a numerical comparison of these factors.

We also compare the reconstruction complexity, in the simpler scenario that the support is fixed. In this setting, recovery is merely a matrix-vector multiplication with the relevant pseudo-inverse: (14) for the MWC or $\mathbf{s}_\Omega = (\boldsymbol{\Phi} \mathbf{F})_\Omega^\dagger \mathbf{y}$ for the RD, where $\Omega$ indicates the active tones, cf. (6). As before, the size of $\boldsymbol{\Phi}$ results in long delay and huge memory length for collecting the samples. The number of scalar multiplications (Mult.-ops.) for applying the pseudo-inverse reveals again orders of magnitude differences. We expressed the Mult.-ops. per block of samples and scaled them to operations per clock cycle of a 100 MHz DSP processor.

We conclude the table with our estimation of the technology barrier of each approach. Computational loads and memory requirements in the digital domain are the bottleneck of the RD approach. Therefore the size of CS problems that can be solved with available processors limits the recovery. We estimate that a Nyquist-rate of $W \approx 1$ MHz may be already quite demanding using convex solvers, whereas $W \approx 10$ MHz is probably the barrier using greedy methods[2]. The MWC is limited by the technology for generating the periodic waveforms $p_i(t)$, which depends on the specific choice of waveform. The estimated barrier of 23 GHz refers to implementation of the periodic waveforms according to [42], [43], though realizing a full MWC system at these high rates can be a challenging task. Our barrier estimates are roughly consistent with the hardware publications of these system: [46], [47] report the implementation of (single, parallel) RD for Nyquist-rate $W = 800$ kHz. An MWC prototype demonstrates faithful reconstruction of wideband inputs with $f_{\mathrm{NYQ}} = 2$ GHz [28].

### F. Choice of Analog Compression

The comparison between the RD and MWC systems reveals how two seemingly-similar choices of analog preprocessing can result in different performance, in terms of the three metrics we considered: robustness to model mismatch, required hardware accuracy and computational loads. Based on the insights gained, we draw several operative conclusions for the choice of $P$:

1) set system parameters with safeguards to accommodate possible model mismatches,
2) incorporate design constraints on $P$ that suit the technology generating the source signals, and
3) balance between nonlinear (subspace detection) and linear (interpolation) reconstruction complexities.

The first point follows immediately from Fig. 5 and basically implies that model and sampler parameters should not be tightly related, implicitly or explicitly. We elaborate below on the other two suggestions.

Input signals are eventually generated by some source, which has its own accuracy specifications. Therefore, if designing $P$ imposes constraints on the hardware that are not stricter than those required to generate the input signal, then there are no essential limitations on the input range. We support this conclusion by several examples. The MWC requires accuracy that is achieved with RF technology, which also defines the possible range of multiband transmissions. The same principle of shifting spectral slices to the origin with different weights can be achieved by periodic nonuniform sampling [16]. This strategy, however, can result in a narrower input range that can be treated, since current RF technology can generate source signals at frequencies that exceed front-end bandwidths of existing ADC devices [8]. Multiband inputs generated by optical sources, however, may require a different compression stage $P$ than that of the RF-based MWC system.

---

[2]A bank of RD channels was studied in [46]. The parallel system duplicates the analog issues and its computational complexity is not improved by much.

**TABLE IV:** Discretization Impact on Computational Loads

| | **RD** | | | **MWC** | |
|---|---|---|---|---|---|
| | Discretization spacing | $\Delta = 1$ Hz | $\Delta = 100$ Hz | | |
| Model | $K$ tones | $300 \cdot 10^6$ | $3 \cdot 10^6$ | $N$ bands | 6 |
| | out of $Q$ tones | $10 \cdot 10^9$ | $10 \cdot 10^7$ | width $B$ | 50 MHz |
| Sampling setup | alternation speed $W$ | 10 GHz | 10 GHz | $m$ channels[§] | 35 |
| | | | | $M$ Fourier coefficients | 195 |
| | rate $R$, eq. (9), theory | 2.9 GHz | 2.9 GHz | $f_s$ per channel | 51 MHz |
| | eq. (7), practice | 5 GHz | 5 GHz | total rate | 1.8 GHz |
| Underdetermined system | (8): $\mathbf{y} = \mathbf{HDFs}$, $\|\mathbf{s}\|_0 \leq K$ | | | (13): $\mathbf{V} = \mathbf{CU}$, $\|\mathbf{U}\|_0 \leq 2N$ | |
| Preparation | | | | | |
| Collect samples | Num. of samples $N_R$ | $5 \cdot 10^9$ | $5 \cdot 10^7$ | $2N$ snapshots of $\mathbf{y}[n]$ | $12 \cdot 35 = 420$ |
| Delay | $N_R/R$ | 1 sec | 10msec | $2N/f_s$ | 235nsec |
| Complexity | | | | | |
| Matrix dimensions | $\mathbf{\Phi} = \mathbf{HDF} = N_R \times Q$ | $5 \cdot 10^9 \times 10^{10}$ | $5 \cdot 10^7 \times 10^8$ | $\mathbf{C} = m \times M$ | $35 \times 195$ |
| Apply matrix[♯] | $\mathcal{O}(W \log W)$ | | | $\mathcal{O}(mM)$ | |
| Storage[♯] | $\mathcal{O}(W)$ | | | $\mathcal{O}(mM)$ | |
| Real time (fixed support) | $\mathbf{s}_\Omega = (\mathbf{\Phi F})^\dagger_\Omega \mathbf{y}$ | | | (14): $\mathbf{z}_\lambda[n] = \mathbf{C}^\dagger_\lambda \mathbf{y}[n]$ | |
| Memory length | $N_R$ | $5 \cdot 10^9$ | $5 \cdot 10^7$ | 1 snapshot of $\mathbf{y}[n]$ | 35 |
| Delay | $N_R/R$ | 1 sec | 10msec | $1/f_s$ | 19.5nsec |
| Mult.-ops. (per window) | $KN_R$ | $1.5 \cdot 10^{18}$ | $1.5 \cdot 10^{14}$ | $2Nm$ | 420 |
| (100 MHz cycle) | $KN_R/((N_R/R) \cdot 100\text{M})$ | $1.5 \cdot 10^{10}$ | $1.5 \cdot 10^6$ | $2Nmf_s/100\text{M}$ | 214 |
| Reconstruction | 1 DAC at rate $W = 10$ GHz | | | $N = 6$ DACs at individual rates $f_s = 51$ MHz | |
| Technology barrier (estimated) | CS algorithms ($\sim$10 MHz) | | | Waveform generator ($\sim$23 GHz) | |

[§] with $q = 1$; in practice, hardware size is collapsed with $q > 1$ [28].    [♯] for the RD, taking into account the structure $\mathbf{HDF}$.

Along the same line, time-domain accuracy constraints may limit the range of multitone inputs that can be treated in the RD approach, if these signals are generated by RF sources. On the other hand, consider a model of piecewise constant inputs, say with knots at the integers and only $K$ nonidentically-zero pieces out of $Q$. Sampling these signals with the RD system would map to (8), but with an identity basis instead of the DFT matrix $\mathbf{F}$. In this setting, the time-domain accuracy required to ensure that the mapping to (8) holds is within the tolerances of the input source.

Moving on to our third suggestion, we attempt to reason the computational loads encountered in Table IV. Over 1 second, both approaches reconstruct their inputs from a comparable set of numbers; $K = 300 \cdot 10^6$ tone coefficients or $2Nf_s = 612 \cdot 10^6$ amplitudes of active sequences $z_l[n]$. The difference is, however, that the RD recovers all these unknowns by a single execution of a nonlinear CS algorithm on the system (8), which has large dimensions. In contrast, the MWC splits the recovery task to a small-size nonlinear part (*i.e.*, CTF) and real-time linear interpolation. This distinction can be traced back to model assumptions. The nonlinear part of a multitone model, namely the number of subspaces $|\Lambda| = \binom{Q}{K}$, is exponentially larger than $\binom{M}{2N}$ which specifies a multiband union of the same Nyquist bandwidth. Clearly, a prerequisite for balancing computation loads is an input model with as many unknowns as possible in its linear part (subspaces $\mathcal{A}_\lambda$), so as to decrease the nonlinear cardinality $|\Lambda|$ of the union. The important point is that in order to benefit from such modeling, $P$ must be properly designed to incorporate this structure and reduce computational loads.

For example, consider a block-sparse multitone model with $K$ out of $Q$ tones, such that the active tones are clustered in $K/d$ blocks of length $d$. A plain RD system which does not incorporate this block structure would still result in a large $R \times W$ sensing matrix with its associated digital complexities. Block-sparse recovery algorithms, *e.g.*, [48], can be used to partially decrease the complexity, but the bottleneck remains the fact that the hardware compression is mapped to a large sensing matrix[3]. A potential analog compression for this block-sparse model can be an MWC system designed for $N = K/d$ and $B = d\Delta$ specifications.

Our conclusions here stem from the study of the RD and MWC systems, and are therefore mainly relevant for choosing $P$ in Xampling systems that maps their hardware to underdetermined systems and incorporate CS algorithms for recovery. Nonetheless, our suggestions above do not necessitate such a relation to CS, and may hold more generally with regard to other compression techniques.

## IV. X-DSP: Sub-Nyquist Signal Processing

In this section, we study the X-DSP stage of Fig. 1, which targets lowrate DSP. Whilst the streaming measurements enter the digital domain at a low rate, they often cannot be used directly for DSP purposes. For example, the MWC sequences $y_i[n]$ contain a mixture of information bands, whereas standard DSP algorithms expect treating an individual band at a time,

---

[3]Note that simply modifying the chipping and integrate-dumping intervals, in the existing scheme of Fig. 2, to $d$ times larger results in a sensing matrix smaller by the same factor, though (8) in this setting would force reconstructing each block of tones by a single tone, presumably corresponding to a model of $K/d$ active tones out of $Q/d$ at spacing $d\Delta$.

as provided to them by RF demodulation when the carrier frequencies $f_i$ are known.

More broadly, the difficulty in directly processing the X-ADC output stems from the fact that popular DSP algorithms assume an input stream at the Nyquist rate. A fundamental reason for processing at the Nyquist rate is the clear relation between the spectrum of $x(t)$ and that of its pointwise values $x(nT)$, so that digital operations can be easily substituted for their continuous counterparts. Digital filtering is an example where this relation is successfully exploited. Since the power spectral densities of continuous and discrete random processes are associated in a similar manner, estimation and detection of parameters of analog signals can be performed by DSP. When sampling below Nyquist, this key relation no longer holds in general. As before, we study X-DSP by gaining insights from DSP options available in the RD and MWC systems, and later on generalize these insights to broader conclusions for an arbitrary X-DSP stage.

### A. Coarse and Fine Subspace Detection

We begin by considering multiband signals and defining the lowrate DSP goal we would like to achieve. A multiband signal can be described in quadrature representation as [49]:

$$x(t) = \sum_{i=1}^{N/2} I_i(t)\cos(2\pi f_i t) + Q_i(t)\sin(2\pi f_i t), \qquad (21)$$

where $I_i(t), Q_i(t)$ are real-valued narrowband signals, and $f_i$ are relatively high carrier frequencies. Classic communication methods obey (21), including analog amplitude-, phase- and frequency-modulation (AM/PM/FM). Modern digital communication transmit bits using techniques, such as frequency- and phase-shift keying (FSK/PSK), which also conform with (21). In all these communication techniques, the message of interest is encoded in $I_i(t), Q_i(t)$, which are therefore referred to as the information signals. The carrier $f_i$ itself does not contain signal information. When the carrier frequencies $f_i$ of a multiband input are known, the receiver demodulates the carrier frequency $f_i$ and obtains $I_i(t), Q_i(t)$, which are then sampled at a low rate. DSP takes place from that point. In the union settings, our goal is therefore to provide the same samples of $I_i(t), Q_i(t)$, despite the lack of information on the carriers $f_i$. The important point is to obtain $I_i(t), Q_i(t)$ with computational complexity that is proportional to $NB$, without resorting to Nyquist-rate computations or interpolations. For simplicity, in this section $1/T = B$ is assumed, so that the width of a spectrum slice is equal to the (maximal) width of an individual band.

The CTF block in the MWC system performs subspace detection by finding the input spectral support at the coarse resolution of active spectrum slices. This coarse resolution is used in order to meet the design metrics discussed earlier and is sufficient for reconstruction purposes via (14). For DSP purposes, however, a coarse subspace detection is insufficient, since the information signals $I_i(t), Q_i(t)$ are not organized in $z_l[n]$ as standard DSP algorithms expect to receive. For example: in Fig. 4, the energy of the $i$th band splits between two consecutive sequences $z_{l-1}[n], z_l[n]$. A single slice may, in general, contain several information bands. Moreover, even when $z_l[n]$ contains a single band, conventional software does not accommodate the lack of a nominal value for the carrier $f_i$. The fact that $f_i$ is somewhere within a slice width, *e.g.*, a range of $1/T = 51$ MHz in the example of Table IV, does not help, since standard software packages can tolerate only slight offsets from the nominal $f_i$; those that presumably occur due to slight frequency shifts between the transmitter and receiver oscillators. What we need is a fine subspace detection, at the level of the union model of (21), in which a fine subspace is defined by $\lambda_{\text{fine}} = \{f_i\}$ and each $\mathcal{A}_{\lambda_{\text{fine}}}$ contains the corresponding information signals $I_i(t), Q_i(t)$.

The algorithm we develop in the sequel refines the subspace detection and outputs an accurate estimate of $f_i$ and samples of the pair $I_i(t), Q_i(t)$, per each band $1 \leq i \leq N/2$, thereby enabling processing at baseband rates with conventional DSP algorithms. For the development, we need to assume that $I(t), Q(t)$ are random with zero cross-correlation, $\mathbb{E}[I(t_1)Q(t_2)] = 0$ for all $t_1, t_2$. In practice, this means that $I(t), Q(t)$ carry uncorrelated information messages. This holds for AM, by definition, and for many digital communication techniques, when using a preceding source coding stage [49]. The algorithm does not assume any specific modulation technique; the only essential assumption is the quadrature form (21) and zero cross-correlation between $I(t), Q(t)$. We refer to the proposed algorithm as Back-DSP.

### B. Algorithm Description

The Back-DSP algorithm consists of three steps:

1) Refining the coarse support estimate $\lambda$ to the actual band edges $[a_i, b_i]$. Here, we rely on two additional model parameters: the minimal width of a single band $B_{\text{min}}$ and the smallest spacing between bands $\Delta_{\text{min}}$. These quantities are often known in communication, though uncertainty in the values $B_{\text{min}}, \Delta_{\text{min}}$ has little effect on the performance, as described later on;
2) Generating $s_i[n]$ per band $1 \leq i \leq N/2$. This step processes $z_l[n]$ and incorporates the edges $[a_i, b_i]$; and
3) Estimating $f_i$ using a digital version of the balanced quadricorrelator (BQ) [49].

The information signals $I_i(t), Q_i(t)$ are obtained upon completion at no additional cost.

Algorithm 1 outlines the operations that are carried out in each step of Back-DSP, whose technical steps are expanded below. We specify the MATLAB commands (in verbatim font) that are used in our implementation. A software package of the Back-DSP algorithm is available online in [50].

**Step 1.** For convenience, the complex-valued $z_l[n]$ are converted to real-valued counterparts $x_l[n]$, taking into account the conjugate-symmetry of $x(t)$. The sequence $x_l[n]$ is obtained by re-positioning $z_l[n], z_{-l}[n]$ on both sides of the origin. Mathematically, $x_l[n] = I_{2,0.5B}\{z_{\pm l}[n]\}$, where

$$I_{r,F}\{z_{\pm l}[n]\} \overset{\triangle}{=} (z_l[n] \uparrow r)e^{-j2\pi Fn} + (z_{-l}[n] \uparrow r)e^{j2\pi Fn}, \qquad (22)$$

and $\uparrow r$ denotes rate increase by a factor of $r$, with the appropriate post-filtering (interpft). By abuse of notation,

---

**Algorithm 1:** Backward-compatible DSP (Back-DSP)

---

**Step 1: Band edges estimation** $[a_i, b_i]$

1.1   for each $l \in \lambda$
$$x_l[n] = \begin{cases} I_{2,0.5B}\{z_{\pm l}[n]\} & l \neq 0 \\ z_0[n] & l = 0 \end{cases}$$

1.2   for each $l \in \lambda, l \neq 0$
      Estimate PSD of $x_l[n]$
      Threshold (24) → fine support estimate

1.3   unite too adjacent intervals, $\leq \Delta_{\min}$ (*combat noise*)
      prune too narrow intervals, $\leq B_{\min}$ (*false alarms*)
      retain only $N/2$ "powerful" intervals (*model assumption*)

**Step 2: Isolate sequence** $s_i[n]$ **per band,** $i = 1, \ldots, N/2$:

2.1   Stitch bands energy
$$\tilde{s}_i[n] = \begin{cases} x_l[n] & \text{same slice} \\ I_{4,0.5B}\{z_{\pm l}[n]\} + I_{4,B}\{z_{\pm(l+1)}[n]\} & \text{band split} \end{cases}$$

2.2   Filter out-of-band contents → $s_i[n]$

**Step 3: Carrier estimate** $f_i, i = 1, \ldots, N/2$:

3.1   Upsample $\uparrow 3$ and frequency-shift

3.2   Apply BQ, Fig. 8

---

here and in the sequel the same index $n$ is used before and after the rate conversion, where the context resolves the ambiguity. The case $l = 0 \in S$, has $x_0[n] = z_0[n]$.

To find the band edges $[a_i, b_i]$, we estimate the power spectral density (PSD) of $x_l[n]$. We used the Welch PSD estimation method [51] (pwelch), with a windows overlapping ratio of 50%, and the shortest window length $W_{\text{size}}$ that meets the frequency resolution

$$f_{\text{res}} = \min(B_{\min}, \Delta_{\min}), \quad W_{\text{size}} \geq \frac{2B}{f_{\text{res}}}. \qquad (23)$$

The PSD estimation produces $P_{xx}^{(l)}[k]$ for $1 \leq k \leq K \approx W_{\text{size}}/2$. A logarithmic threshold

$$\log_{10}(\text{Threshold}) = \frac{1}{K} \sum_{k=1}^{K} \log_{10} P_{xx}^{(l)}[k], \qquad (24)$$

translates $P_{xx}^{(l)}[k]$ to a binary decision on the energy concentration.

To mitigate undesired noise effects; support regions that are closer than $\Delta_{\min}$ are united, and isolated regions with widths smaller than $B_{\min}$ are pruned. Our final estimate of the band edges $[a_i, b_i]$ comprises the $N/2$ most powerful bands, according to the PSD values.

**Step 2.** The purpose of this step is to obtain a sequence $s_i[n]$ for each $1 \leq i \leq N/2$, such that $s_i[n]$ contains the entire contribution of exactly one band. Using the edges $[a_i, b_i]$ we identify the cases of band split, namely when the energy of $s_i(t)$ resides in adjacent spectrum slices $x_l[n], x_{l+1}[n]$ for some $0 \leq l \in \lambda$; see Fig. 4 for example. In such cases, merging occurs via

$$\tilde{s}_i[n] = I_{4,0.5B}\{z_{\pm l}[n]\} + I_{4,B}\{z_{\pm(l+1)}[n]\}, \qquad (25)$$

otherwise, $\tilde{s}_i[n] = x_l[n]$ for a frequency band $[a_i, b_i]$ that lies in a single spectrum slice. As a result, $\tilde{s}_i[n]$ contains the entire energy of the $i$th band, possibly with additional contributions due to other information bands. We use the estimated edges $[a_i, b_i]$ to filter $\tilde{s}_i[n]$ from the out of band contents (firpmord and firpm are used in our implementation [50]). The filter type can be either low-, high-, band- or all-pass, depending on the locations of the other bands, as illustrated in Fig. 7. The allowed ripples in the pass- and stop-bands are $A_p = 10^{-6}, A_s = 10^{-2}$, respectively. The filter order is often small, since the actual spacing between the bands relaxes the cutoff constraints.

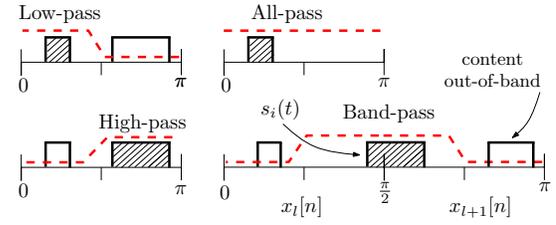

**Fig. 7:** Filtering out-of-band noise in Step 2 of algorithm Back-DSP.

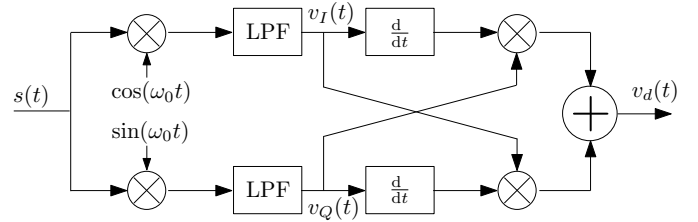

**Fig. 8:** The analog balanced-quadricorrelator [49].

**Step 3.** The final step estimates the carriers $f_i$. Rough estimates can be readily computed at the median frequencies $(a_i + b_i)/2$, though we observed that this estimate is inaccurate in noisy settings. To improve the estimate, we use the BQ whose circuit appears in Fig. 8. For brevity, the band index $i$ is omitted. The BQ is an analog circuit that estimates the carrier frequency of a quadrature input $s(t)$ that consists of a single pair of information signals. It is initialized with an angular frequency $\omega_0 = 2\pi f_0$ and outputs $v_d(t)$ whose expected value is proportional to offset from the true carrier $f_c$

$$\mathbb{E}[v_d(t)] = -K_G(f_c - f_0)(\mathbb{E}[I^2(t)] + \mathbb{E}[Q^2(t)]). \qquad (26)$$

In practice, time averaging replaces the expectations. The constant $K_G$ in (26) is the effective analog gain of the mixers, filters and differentiators along the way.

In our algorithm, we implement a digital version of the BQ and used FIR lowpass filters and approximated the continuous derivatives by the finite difference – a filter with the discrete impulse response $[1, -1]$. Note that a wide family of filters can substitute the true differentiators [49].

A fundamental requirement of the BQ, either in analog or digital, is that the first mixing yields non-overlapping copies of $s(t)$ at $\omega_0 \pm \omega_c$. To ensure this property, each $s_i[n]$ is interpolated by a factor of three, and the positive and negative frequencies are re-positioned in angular positions $[\pi/3, 2\pi/3], [-2\pi/3, -\pi/3]$, respectively. For example, when no merging occurs in Step 2.1, this computation boils down to $I_{6,1.5B}\{z_{\pm l}[n]\}$ with the relevant $l$. The digital BQ is initialized to the angular frequency matching $(a_i + b_i)/2$ and applied iteratively. Each iteration refines the previous estimate

by

$$\omega_0^{\text{new}} = \omega_0^{\text{old}} + G \frac{\sum_n v_d[n]}{\sum_n |s_i[n]|^2}, \qquad (27)$$

with a loop gain $G = 5 \cdot 10^6$. The procedure monitors $\omega_0 \in [\pi/3, 2\pi/3]$ and terminates upon convergence or if a pre-defined number of iterations is reached.

**Properties.** Upon completion, the (samples of the) desired information signals $I_i(t), Q_i(t)$ of the $i$th band are readily available – the last BQ iteration computed them for the nodes $v_I(t), v_Q(t)$ of Fig. 8. The rate of $I_i[n], Q_i[n]$ is either $6B$ or $12B$, depending on the rate of $s_i[n]$. The recovered carrier $f_i$ and the detected band edges $[a_i, b_i]$ allow to reduce the rate of $I_i[n], Q_i[n]$ to minimum, *i.e.*, $2(b_i - a_i)$.

Besides the information signals $I(t), Q(t)$, the algorithm outputs additional useful information per band: the edges $[a_i, b_i]$, the isolated sequence $s_i[n]$ and the carrier estimate $\hat{f}_i$. The latter is computed from the angular frequency $\omega_0$ that the BQ converged to as

$$\hat{f}_i = B \left( l + c \frac{\omega_0 - \pi/3}{\pi/3} \right), \qquad (28)$$

where $c = 1$ when merging was not required, and $c = 2$ otherwise. The carrier-frequency-offset (CFO) $\hat{f}_i - f_i$ is not expected to be zero, but rather to fall below the allowed tolerance of commercial standards, as if a nominal $f_i$ value was specified. We report the actual CFO values in the next subsection.

For applications in which the exact $B_{\min}, \Delta_{\max}$ are unknown, approximate values can be set. The uncertainty with respect to the true values may yield many possible support regions in steps (1.1)-(1.2). Nonetheless, the effect on the overall performance is minor, since only the $N/2$ powerful regions are selected in step 1.3. The exact band locations have only a negligible effect on the filter design in step 2.2. Furthermore, the BQ in step 3 is insensitive to inaccuracies in $[a_i, b_i]$. Therefore, approximate values for $B_{\min}, \Delta_{\max}$ are sufficient in practice. We used $B_{\min} = \Delta_{\max} = B/8$ in our simulations.

As a nice feature, using the proposed algorithm, the original continuous reconstruction of Fig. 3 can now be improved. In [8], $x(t)$ is reconstructed from $z_l[n]$ by interpolation and properly positioning the spectrum slices. Since the scenario of band splitting can be fairly common, at most $2N$ spectrum slices may be active, hence the number of DAC and modulation branches. With Back-DSP, we can now reconstruct $x(t)$ using (21), which requires only $N$ mixers, filters and DACs. In addition, note that once the information signals $I_i(t), Q_i(t)$ are obtained, error correction DSP algorithms can be employed to improve the overall robustness to noise.

### C. Simulations

To evaluate the accuracy of the estimate $\hat{f}_i$, we simulated an example multiband model with $N = 6, B = 50$ MHz. Quadrature phase-shift keying (QPSK) modulation was used

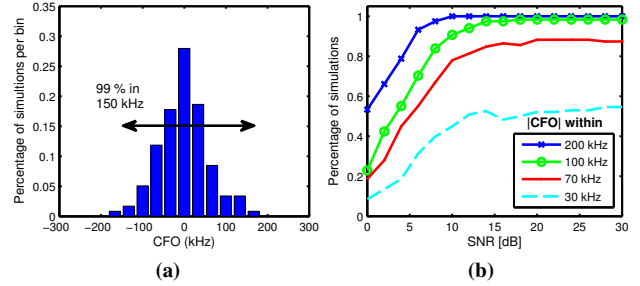

**Fig. 9:** The distribution of CFO for fixed SNR=10 dB (a). The curves (b) represent the percentage of simulations in which the CFO magnitude is within the specified range.

to generate $x(t) = \sum_{i=1}^{3} x_i(t)$ via

$$x_i(t) = \sqrt{\frac{2E_i}{T_{\text{sym}}}} \left( \sum_n I_i[n] p(t - nT_{\text{sym}}) \right) \cos(2\pi f_i t) \quad (29)$$

$$+ \left( \sum_n Q_i[n] p(t - nT_{\text{sym}}) \right) \sin(2\pi f_i t) + n(t),$$

where $E_i = \{1, 2, 3\}$, $1/T_{\text{sym}} = 30$ MHz, $p(t) = \texttt{rcosine}(t/T_{\text{sym}})$ are the symbol energy, rate and raised-cosine pulse shape with 30% rolloff, respectively. The carriers $f_i \in [0, 5]$ GHz, the bit streams $I_i[n] = \pm 1, Q_i[n] = \pm 1$, and the additive white Gaussian noise $n(t)$ were all drawn independently at random.

An MWC with the basic configuration (10) was used with $m = 30$ channels and sign alternating waveforms $p_i(t)$, $M = 195$ alternations points per period $T$. We assume the spectrum slices $z_l[n]$ were obtained successfully by the preceding stages of Fig. 3. For each one of 40 test signals, we executed the Back-DSP algorithm and measured the CFO $\hat{f}_i - f_i$ for $i = 1, 2, 3$. Fig. 9 reports the distribution of the CFOs encountered in our simulations for various signal-to-noise ratios (SNRs). Evidently, in most cases our algorithm approaches the true carriers as close as 150 kHz. For reference, the 40ppm CFO specifications of IEEE 802.11 standards tolerate 150 kHz offsets for transmissions located around 3.75 GHz [52].

To verify data retrieval using the Back-DSP algorithm, we generated a single binary phase-shift keying (BPSK) transmission, such that the band energy splits between two adjacent spectrum slices. We executed the algorithm and used a Costas-loop receiver [53] to extract the bits encoded in the BPSK transmission. We measured the bit error rate (BER), that is the number of erroneous bits at the output, in a Monte Carlo simulation. For each trial out of 2500, we redraw a carrier position that gives band split, and simulated 6000 bits passing through the analog sampler and the digital algorithms. We repeated the procedure for input SNRs of 3,5,7 and 9 dB. In total, about 15 million bits were simulated. Estimated BERs for 3 dB and 5 dB SNR, respectively, are better than $0.77 \cdot 10^{-6}$ and $0.71 \cdot 10^{-6}$. No erroneous bits were detected for SNR of 7 and 9 dB. Lab experiments in [28] report correct continuous reconstruction of a mixture of AM and FM signals, whose energy overlays at baseband.

### D. X-DSP and Related Work

The Back-DSP algorithm provides the MWC with a smooth interface to existing DSP packages. This backward-compatibility is achieved due to a simple relation between the contents of spectrum slices and the desired information signals $I_i(t), Q_i(t)$. Thus, a coarse subspace detection is sufficient for reconstruction purposes while a finer subspace detection enables lowrate DSP with existing algorithms. In other applications, the same detection algorithm may work for both reconstruction and lowrate DSP. For example, in [9] the union model consists of sequences of innovations which potentially carry information. The active sequences are detected by ESPRIT [24] at a fine resolution, which is also used for reconstruction. In [15], a sparse shift-invariant model is assumed and the CTF is used to detect the active shift-invariant subspaces and their contents at once.

Subspace detection essentially inverts the analog compression operator $P$. Thus, lowrate DSP also depends on the chosen $P$. For example, if we were to treat multiband signals in the RD approach, presumably via discretization to a grid of $\Delta$-spaced tones, achieving DSP at low rates could be more difficult. The price would be the large computational loads of Table IV and additional computations on length-$K$ vectors to extract the information signals $I_i(t), Q_i(t)$ from the recovered tones ($K = 300 \cdot 10^6$ or $3 \cdot 10^6$ depending on $\Delta$).

An interesting related work is [25]. The approach, termed compressive signal processing (CSP), considers the basic CS setup of an underdetermined system $\mathbf{y} = \mathbf{\Phi}\mathbf{x}$ and questions whether the CS measurements $\mathbf{y}$ could be used to infer quantities of interest, without first recovering $\mathbf{x}$. It is shown that certain quantities that are invariant under the sensing operator $\mathbf{\Phi}$ can be determined from $\mathbf{y}$ directly, provided that $\mathbf{\Phi}$ satisfies certain embedding conditions [25]. For example, Euclidean distances are approximately preserved under an underdetermined mapping. Whilst computational-complexity and implementation issues were not concerned in [25], here we attempt to examine the potential of CSP to provide lowrate processing from a set of RD measurements. Since CSP avoids reconstruction, one can, in principle, apply CSP on a small set of RD measurements and thus hope to escape the high computational loads involved in detecting the active tones subspace out of $\binom{Q}{K}$.

In practice, however, high computational loads are not alleviated by the combination CSP-RD, since the stable embedding conditions of CSP require $\mathbf{\Phi}$ to behave as a near isometry on Nyquist-rate sparse vectors, effectively requiring $\mathbf{\Phi}$ with dimensions that can work for reconstruction purposes. This means that one should collect samples until $\mathbf{y}$ corresponds to $\mathbf{\Phi}$ with a sufficiently large number of rows. In turn, as Table IV shows, processing of $\mathbf{y}$ requires computations on vector lengths of $N_R = 5 \cdot 10^9$ or $N_R = 5 \cdot 10^7$ entries, depending on the discretization spacing $\Delta$. Thus, while avoiding the reconstruction of $\mathbf{x}$ which has even higher dimensions, CSP does not lead to substantial savings in this case.

More inherently, CSP calls for the development of a new toolbox of processing methods. The examples in [25] are limited to certain linear processing tasks, whose true values are approximated in the compressed domain. Moreover, CSP algorithms, as those proposed in [25], involve the exact values of the sensing matrix, so that any hardware inaccuracy that alters $\mathbf{\Phi}$ can propagate errors to the CSP algorithms. In contrast, Xampling proposes first to detect the signal subspace and then perform conventional subspace DSP. In essence, Xampling suggests that the DSP does not need to be aware of the source of its input. For example, for multiband transmissions, the data can either arrive from a demodulator that knows the carriers $f_i$ or from the MWC which does not incorporate such knowledge; The information signals $I_i(t), Q_i(t)$ are given the same treatment either case. In practice, both CSP and X-DSP can be relevant, depending on the application at hand.

## V. Concluding remarks

Union of subspaces models appear nowadays at the research frontier in sampling theory. The ultimate goal is to build a complete sampling theory for UoS models of the general form (1) and then derive specific sampling solutions for applications of interest. Although several promising advances have already been made [5]–[7], [15], this esteemed goal is yet to be accomplished.

In this paper, we proposed a unified framework for treating UoS signals from a functional viewpoint. As Table II shows, the proposed Xampling architecture is broad enough to capture a multitude of engineering solutions, under the same logical flow of operations. The core contributions which assisted in developing Xampling are two. First, we examined analog compression through the way it is realized in the RD and MWC systems. Our technical comparison revealed that the somewhat visual resemblance can be quite misleading. Major differences were found in three metrics of practical interest: robustness to model mismatch, hardware complexity and computational loads, with the MWC outperforming in all three aspects for our signals of interest. Based on this study, we have drawn operative conclusions for the choice of analog compression operator in Xampling systems, and in particular for those systems that rely on CS ideas. Second, we addressed the challenge of lowrate DSP and developed the Back-DSP algorithm, which completes the X-DSP functionality for the MWC system, with lowrate processing options via a smooth interface to standard DSP packages.

The nomenclature Xampling was chosen to highlight the important aspects of our framework. The X prefix symbolically represents the intersection between subspaces in a union, so as to highlight that sampling a union model requires a systematically different treatment in acquisition and processing due to the multiple subspaces, yet that this is still a sub-field of generalized sampling theory [1], [2], [54], [55]. Xampling, literally pronounced as CS-Sampling (phonetically /kˈsæmplɪŋ/), also symbolizes a synergy between recent and classic paradigms in sampling, thereby conveying a balance between CS techniques and traditional concepts from sampling theory, and between nonlinear and linear reconstruction techniques. Finally, it was recently suggested to us [56] that X can stand for "extreme sampling", hinting at the very low rates.

## Acknowledgment

The authors would like to thank the anonymous reviewers for their constructive comments and insightful suggestions.